\documentclass[twocolumn,showpacs,preprintnumbers,amsmath,amssymb]{revtex4}

\usepackage{graphicx}
\usepackage{bm}

\begin{document}


\title{Josephson $\pi$-state in superconductor-Luttinger liquid hybrid systems}

\author{Nobuhiko Yokoshi}
\author{Susumu Kurihara}%
\affiliation{%
Department of Physics, Waseda University, Okubo, Shinjuku, Tokyo 169-8555, Japan }%

\date{\today}

\begin{abstract}
Josephson current through a Luttinger liquid (LL) under a magnetic field is theoretically studied. We derive an analytical expression of Josephson current for clean interfaces, by using quasiclassical Green's function and functional bosonization procedure. We show that critical currents can be renormalized by electron-electron interactions at perfect transparency when LL is adiabatically connected with superconductors. We also find that a generation of $\pi$-state, due to spin-dependent energy shift in Andreev bound states (ABS), is prohibited even at zero temperature when the strength of repulsive interactions reaches some critical value. The suppression of $\pi$-state is caused by the low energy fluctuations propagating in LL, and making the Zeeman splitting in ABS blurred.
\end{abstract}

\maketitle


Transport phenomena in one-dimensional (1D) structures are strongly affected by electron-electron interactions. Fermi liquid description breaks down due to electron correlation, and systems are believed to behave as Luttinger liquids (LLs), where low-lying excitations are collective modes rather than single-particle excitations. This gives rise to characteristic phenomena such as spin-charge separation and charge fractionalization. These behaviors have been verified experimentally, e.g., in carbon nanotubes (CNTs)~\cite{rf:1,rf:2} and quantum wires on GaAs/AlGaAs heterostructure~\cite{rf:3,rf:4}. Inspired by advances in microfabrication techniques, hybrid structures connecting LL with other conductors have attracted attentions from theoretical and experimental sides as candidates of new mesoscopic electronic devices.

Recently, superconducting proximity effect on CNTs in contact with superconductors was reported in a couple of experiments~\cite{rf:5,rf:6}. There are some previous works about Josephson currents in superconductor-Luttinger liquid (S-LL) hybrid systems~\cite{rf:7,rf:8,rf:9,rf:10,rf:11,rf:12}. In such systems, electron-electron interactions modify the normal and the Andreev reflection (AR) process, and lead to unique power law behaviors of low energy properties such as proximity effect, conductance and local density of states (LDOS). On the other hand, a generation of metastable $\pi$-state has been studied theoretically~\cite{rf:13} and experimentally~\cite{rf:14,rf:15} in superconductor-ferromagnet (S-F) junctions. The $\pi$-state is a result of spin-dependent energy shift in Andreev bound states (ABSs) due to exchange energy in F. Here, we study Josephson current in the presence of the $\pi$-state in S-LL junctions. By applying a magnetic field to the LL region, we can consider that Zeeman energy in LL plays the same role as the exchange energy in F. In this paper, we derive the analytical expression of Josephson current and find that critical currents can be renormalized by the interactions at perfect transparency when LL is adiabatically connected with reservoirs. We also show that the Coulomb interactions affect the generation of the metastable $\pi$-state through the LDOS in 1D structures.

The system under consideration is a long S-LL-S junction where LL is adiabatically connected with s-wave superconductors (see Fig.1). Approximately influence of the interactions in superconductors is neglected, and superconducting order parameters $\Delta e^{\pm i\chi /2}$ is assumed to change abruptly. For LL region, we assume that Fermi velocity is spin-independent $v_{F\uparrow}=v_{F\downarrow}=v_F$. This assumption may be justified as far as the Zeeman energy is far smaller than the Fermi energy. In this case, we can consider that the spin-charge separation in LL does not break down~\cite{rf:16}. Thus, Hamiltonian with usual g-ology is expressed as
\begin{figure}
\includegraphics[scale=0.5]{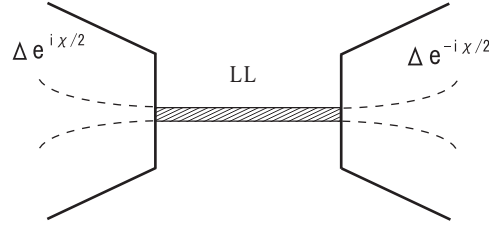}
\caption{\label{fig:epsart}  A possible implementation of system setup: s-wave superconductors are deposited on quantum wire which is adiabatically connected with two dimensional Fermi liquids. We believe that superconducting proximity effect penetrates in Fermi liquid regions. The electrons in superconductors can tunnel into Luttinger liquid via Andreev reflection process. For sake of simplicity, barrier potentials at the junctions are assumed to be negligible.}
\end{figure}
\begin{eqnarray}
\!\!\!\!\!\! H_{\rm LL}= \int dx \Bigl[
&\sum _{a,s}&\psi_{a,s}^{\dagger}
(-i  av_F \frac{\partial}{\partial x}+sh)\psi_{a,s} \nonumber \\
&+&\sum_{s ,s '} g_{2,p} \rho_{+,s}(x,t) \rho_{-,s '}(x,t) \nonumber \\
&+&\sum_{a, s ,s '} \frac{g_{4,p}}{2} \rho_{a,s} (x,t) \rho_{a,s '} (x,t)
\Biggr],
\end{eqnarray}
where $\rho_{a,s}$ and $h$ are particle density and Zeeman energy. Here, $a=\pm$ and $s=\pm$ denote direction of movement and spin, respectively. $p=\{ \parallel, \perp \}$ specifies interaction between electrons with parallel or anti-parallel spins. We consider the case where back-scattering and umklapp-scattering processes are irrelevant for simplicity. Throughout this paper, $\hbar$ and $k_B$ are set to unity. 

We apply a so-called Hubbard-Stratonovich transformation to the interaction terms~\cite{rf:17}, and obtain the action using auxiliary field $\phi_{a,s}(x,t)$ as follows
\begin{eqnarray}
\!\!\!\! S[\phi]=\int dtdx \Bigl[
&L_0(\psi^{\dagger},\psi)&+L_1(\phi) \nonumber \\
&+&\sum _{a,s} \phi_{a,s}(x,t) \rho_{a,s}(x,t)
\Bigr].
\end{eqnarray}
Here, $L_0$ and $L_1$ are the Lagrangian densities of free fermions and collective fluctuations; they can be expressed in matrix form as
\begin{eqnarray}
L_0&=&[ \psi,(i\frac{\partial}{\partial t}\hat{1}+ iv_F \frac{\partial}{\partial x}\hat{\tau}_z \hat{\Sigma}_z+h\hat{\tau}_z)\psi ], \\ 
L_1&=& [\phi,\hat{\rm g}^{-1}\phi ],
\end{eqnarray}
where $\hat{\rm g}$ represents interactions, and
\begin{eqnarray}
\hat{\tau}_i= 
\left(
\begin{array}{cc}
{\boldsymbol \sigma}_i & {\bf 0} \\ {\bf 0} & {\boldsymbol \sigma}_i
\end{array}
\right), \ \ \ \ 
\hat{\Sigma}_z= 
\left(
\begin{array}{cc}
{\bf 1} & {\bf 0} \\ {\bf 0} & {\bf -1}
\end{array}
\right)
\end{eqnarray}
with ${\boldsymbol \sigma}_i$ being usual Pauli matrices. Throughout the paper, quantities with gcareth denotes $(4 \times 4)$ matrices, and those with boldface $(2 \times 2)$ matrices. 1st and 3rd row correspond to right and left moving electrons with spin up, and 2nd and 4th row to left and right moving holes with spin down. From Eq.(2), one can regard LL as the free fermions propagating in the bosonic ``environment''~\cite{rf:18}. Here the free fermion part includes the role of topological term in usual bosonization method. Therefore, we treat $\phi(x,t)$ as ``local scalar potential'' for a time, and average it in terms of fluctuation fields. One can obtain quasiclassical Green's function in LL by solving Eilenberger equation~\cite{rf:19}
\begin{eqnarray}
&&iv_F\frac{\partial}{\partial x}\hat{g}(x,t,t'|\phi)+ \nonumber \\
&&\Bigl[
(i  \frac{\partial}{\partial t} +h)
 \hat{\tau}_z \hat{\Sigma }_z 
+\phi(x,t) \hat{\Sigma }_z,\  \hat{g}(x,t,t'|\phi) 
\Bigr]_-=0,
\end{eqnarray}
where $[ \cdots ]_-$ denotes commutator. 

In a similar fashion, one can construct the stationary Green's functions in S regions~\cite{rf:20}. Here, it is assumed that the magnetic field is not applied to S region. In Fourier representation with respect of time difference, they are given by 
\begin{eqnarray}
\hat{g}_s^{L(R)}(x,\epsilon)=\hat{U}(\hat{g}_0(\epsilon)+\hat{g}^{L(R)}(x,\epsilon))\hat{U}^{\dagger},
\end{eqnarray}
where $\hat{U}={\rm exp}[i\chi\hat{\tau}_z/4]$. $\hat{g}_0(\epsilon)=(\epsilon \hat{\tau}_z+i\Delta \hat{\tau}_y)/\Omega(\epsilon)$ is the one at far from the interfaces, and 
\begin{eqnarray}
&&\hat{g}^{L(R)}(x,\epsilon)=\hat{C}^{L(R)}(\epsilon)e^{\mp 2i\Omega(\epsilon)\frac{x\pm L/2}{v_F}} \times \nonumber \\
&&\left( 
\begin{array}{cc}

\begin{array}{cc}
1 & \alpha (\epsilon)^{\mp 1} \\ -\alpha (\epsilon)^{\pm 1} & -1
\end{array} 
& {\bf 0} 
\\ 
{\bf 0} 
& 
\begin{array}{cc}
1 & \alpha (\epsilon)^{\pm 1} \\ -\alpha (\epsilon)^{\mp 1} & -1
\end{array}

\end{array}
\right),
\end{eqnarray}
where L (R) depicts S of left (right) hand side. Here, $\Omega(\epsilon)=\sqrt{\mathstrut{\epsilon^2-\Delta^2}}$ and $\alpha(\epsilon)=\Delta/(\epsilon+\Omega(\epsilon))$. One can see that $\alpha(\epsilon)$ is equal to the AR amplitude for interfaces with non-interacting normal conductor. We determine coefficients $\hat{C}^{L(R)}(\epsilon)$ by applying boundary conditions at the interfaces $x=\pm L/2$. Since we focus on the junction with clean interfaces, the boundary conditions are that they are continuous along a trajectory, i.e., $\hat{g}(\pm \frac{L}{2}-0,t,t'|\phi)=\hat{g}(\pm \frac{L}{2}+0,t,t'|\phi)$~\cite{rf:21}. 

If $L$ is larger than coherence length $\xi=v_F/\pi \Delta$, we find that ABS has the following form
\begin{eqnarray}
E_{a,s}^{(n)}&=&\frac{v_F}{2L}\big[(2n+1)\pi+a\chi -s\gamma -\Phi_{a,s} \big]
\nonumber \\
&=&\epsilon_{n,s}-\frac{v_F}{2L}\big[a\chi+\Phi_{a,s} \big],
\end{eqnarray}
where $n$ is an integer, and $\gamma =2Lh/v_F \ ({\rm mod}\ 2\pi)$ is phase shift due to the magnetic field. Here, the influence of the auxiliary field appears as a small energy shift and is
\begin{eqnarray}
\Phi_{a,s}=\theta_{a,s} \bigl(\frac{L}{2},t \bigr)+\theta_{-a,-s}\bigl(\frac{L}{2},t \bigr) - \{ \frac{L}{2} \rightarrow -\frac{L}{2} \},
\end{eqnarray}
with $\theta$'s being ``gauge'' fields in terms of the fluctuations defined by $(\partial_t+av_F\partial_x)\theta_{a,s}(x,t)=\phi_{a,s}(x,t)$~\cite{rf:18,rf:22}. To evaluate the effect of $\Phi_{a,s}$ to the Josephson current, we have to average the Green's function in terms of auxiliary $\phi$ fields; $\hat{g}(x,t,t')=<\hat{g}(x,t,t'|\phi)>_{\phi}$, where
\begin{eqnarray}
< \dots >_{\phi}=
\frac{\int \prod_{a,s}{\mathcal D}\phi_{a,s}  
\dots  e^{iS_{\rm ind}[\phi]}}{\int \prod_{a,s}{\mathcal D}
\phi_{a,s} e^{iS_{\rm ind}[\phi]}}.
\end{eqnarray}
Here, $S_{\rm ind}=\int dtdx L_1$ is the action for the fluctuations induced by the interactions and forms a gaussian in terms of $\theta$ fields. Considering that spin-charge separation is still valid, $S_{\rm ind}$ has the following form after unitary transformation~\cite{rf:18}
\begin{eqnarray}
S_{\rm ind}=\frac{i}{2}[\theta,\hat{M} \theta], \ \ \ 
\hat{M}=
\left(
\begin{array}{cc}
{\bf M}_{\rho} & {\bf 0} \\ {\bf 0} & {\bf M}_{\sigma}
\end{array}
\right),
\end{eqnarray}
where $j=\{ \rho, \sigma  \}$ specifies charge wave or spin wave fluctuations. One see that $\hat{M}$ corresponds to the inverse of Green's function of the fluctuations. Josephson current with s-wave symmetry is related only to charge bosons of the charge waves (i.e. $\theta_{\rho ,c}=\sum_{a,s}\theta_{a,s}/2$) and phase bosons of the spin waves ($\theta_{\sigma ,p}=-\sum_{a,s}as\theta_{a,s}/2$)~\cite{rf:8}. Thus, it is sufficient to have the knowledge of the propagators of these components. The concrete forms in Fourier representation are
\begin{eqnarray}
M^{-1}_{\rho,c}(k,\omega)=i\pi \Bigl[\frac{v_{\rho}(K_{\rho}^{-1}-2)}{\omega^2-v_{\rho}^2k^2}+D^0_{\rho}(k,\omega) \Bigr],
\end{eqnarray}
\begin{eqnarray}
M^{-1}_{\sigma,p}(k,\omega)=i\pi \Bigl[ \frac{v_{\sigma}(K_{\sigma}-2)}{\omega^2-v_{\sigma}^2k^2}+D^0_{\sigma}(k,\omega) \Bigr],
\end{eqnarray}
where
\begin{eqnarray}
\!\!\!D^0_{j}(k,\omega)&=&(v_{j}-v_{F}) \Bigl[\frac{\omega^2+v_Fv_jk^2}{(\omega^2-v_F^2k^2)(\omega^2-v_j^2k^2)} \Bigr].
\end{eqnarray}
Here, $K_{j}$ and $v_j=u_jv_F$ are usual Luttinger parameter and propagation velocity of $j$ component.

\begin{figure}
\includegraphics[scale=0.39]{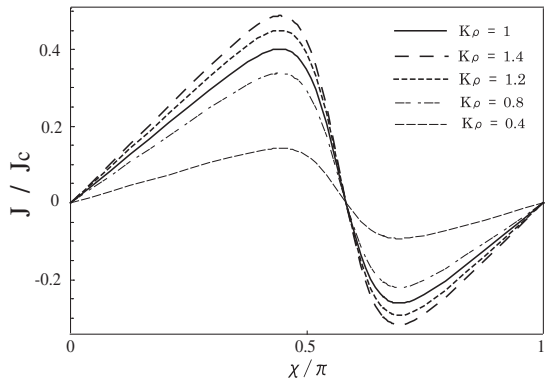}
\caption{\label{fig:epsart}  Josephson current in unit of $J_c=ev_F/L$ is potted as a function of phase difference for different $K_{\rho}$s (Here, we set $\gamma=0.45\pi$, $L/L_T$=0.4, $K_{\sigma}$=1). We assume $g_{i \parallel}=g_{i \perp}$, thus $K_j^{-1}=u_j$. There are positive slopes at $\chi=\pi$ indicating the existence of the $\pi$-state. One can find critical currents are suppressed for repulsive interactions.}
\end{figure}

Total Josephson current through the system can be obtained, summing up the contributions from all the ABSs. Because of the spin-charge separation, renormalization of the spin-up current and the spin-down one is exactly the same. Here we assume that excitation momentum unit of the charge and the spin bosons is infinitesimally small in LL connected with Fermi liquids. As a consequence, Josephson current for S-LL-S system has a following form
\begin{eqnarray}
J=\frac{ev_{F}}{L} \frac{2}{\pi}\frac{L}{L_T}
\sum_{m=1}^{\infty}\sin(m \chi)\cos(m\gamma)
\frac{(-1)^{m+1}\Upsilon^{m^2}}{\sinh(\frac{mL}{L_T})},
\end{eqnarray}
with $L_T=v_F/2\pi T$ being thermal length. The influence of electron-electron interactions is included only in $\Upsilon$, which has a following form
\begin{eqnarray}
\Upsilon =
(\frac{\pi T}
{D})^{K_{\rho}^{-1}+K_{\sigma}-2}
\frac{u_{\rho}^{-K_{\rho}^{-1}}u_{\sigma}^{-K_{\sigma}} 
\sinh ^2(\frac{L}{2L_T})}
{\sinh ^{K_{\rho}^{-1}}(\frac{L}{2u_{\rho}L_T})
\sinh ^{K_{\sigma}}(\frac{L}{2u_{\sigma}L_T})}. \nonumber \\
\end{eqnarray}
Here, $D$ is a high energy cut-off; since we take into account the energy smaller than superconducting energy gap, we assume $D \sim \Delta$~\cite{rf:12}. Eq.(16) shows that each AR process is renormalized by the fluctuations with energies lower than $v_F/L$~\cite{rf:23}. This can be interpreted that the ``emvironment" at the interfaces directly affect the AR. Note that the exponent of renormalization differs from the one due to normal backscattering, e.g., at high barrier potential $\propto (\pi T/D)^{K_{\rho}^{-1}+K_{\sigma}^{-1}-2}$~\cite{rf:24}. Therefore, we consider that it is Cooper pair tunneling process that the interactions renormalize. To see this explicitly, we utilize conventional tunnel Hamiltonian $H_T=\sum_{a,s}\bigr( \mathcal{T}_1\Psi_{a,s}^{\dagger}(-\frac{L}{2})\tilde{\psi}_{a,s}(-\frac{L}{2})+\mathcal{T}_2\Psi_{a,s}^{\dagger}(\frac{L}{2})\tilde{\psi}_{a,s}(\frac{L}{2}) +{\rm H.c.} \bigl)$ with $\Psi_{a,s}(x)$ and $\tilde{\psi}_{a,s}(x)=e^{i(ak_Fx+\theta_{a,s})}$ being annihilation operators in S and LL. At zero temperature, the second order perturbation of $H_T$ at each interface, which corresponds to single AR process, is proportional to
\begin{eqnarray}
&&\int dt<e^{i\bigr( \theta_{+\uparrow}(-\frac{L}{2},0)+\theta_{-\downarrow}(-\frac{L}{2},0)-\theta_{-\downarrow}(\frac{L}{2},t)-\theta_{+\uparrow}(\frac{L}{2},t) \bigl)}>_{\phi} \nonumber \\
&&\propto (\frac{v_F}{DL})^{K_{\rho}^{-1}+K_{\sigma}-2}=\Upsilon(T \rightarrow 0).
\end{eqnarray}
This is in agreement with the result of Fazio {\it et al.}~\cite{rf:7}. One can verify that higher order contributions are consistent with Eq.(16) as far as the power law  dependence is concerned~\cite{rf:18}.

\begin{figure}
\includegraphics[scale=0.4]{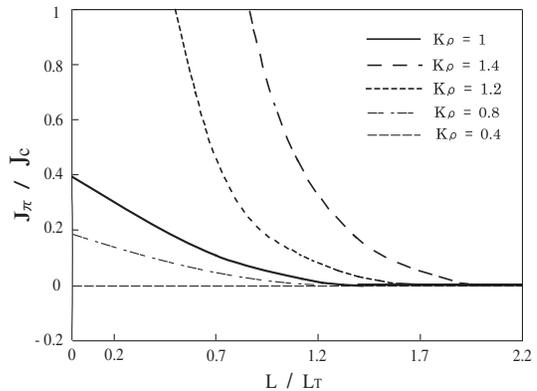}
\caption{\label{fig:epsart} Temperature dependence of $J_{\pi}$ in unit of $J_c$ is plotted at $\gamma=0.35\pi$ and $K_{\sigma}=1$. For repulsive interactions, one can see that $\pi$-state is strongly suppressed. Further, at $K_{\rho}=0.4$, the generation of $\pi$-state is entirely suppressed even at zero temperature.}
\end{figure}

Current-phase relations for different values of $K_{\rho}$ are shown in Fig 2 (Hereafter, we employ the parameters, such as $\Delta$ and $L$, modeling the experiment by Kasumov {\it et al.}~\cite{rf:5}). Owing to the Zeeman phase shift $\gamma$, the positive slopes are found to appear at $\chi=\pi$ ($\pi$-state). This is the same result as in S-F junction except the suppression (enhancement) by the repulsive (attractive) interactions~\cite{rf:13}. However, the suppression by the repulsive interactions ($K_\rho<1$) can play a key role as for the generation of $\pi$-state. We calculate the temperature dependence of the sub-critical current $J_{\pi}$ which provides an indication of the existence of the $\pi$-state (see Fig. 3). In presence of the repulsive interactions, there is the regime for finite $\gamma$ where the $\pi$-state does not occur even at absolute zero, although finite Josephson current can flow. The condition for the existence of a $\pi$-state is given by 
\begin{eqnarray}
\Lambda =-\sum_{m=1}^{\infty}\cos (m \gamma)(\frac{v_F}{DL})^{(K_{\rho}^{-1}+K_{\sigma}-2)m^2} >\ 0 \ .
\end{eqnarray}
In non-interacting case ($K_{\rho}=K_{\sigma}=1$), $\Lambda=0.5$ for finite $\gamma$. In contrast, if the Coulomb interactions are strong enough ($K_{\rho} \lesssim 0.2$) as in CNTs~\cite{rf:2}, the metastable $\pi$-state cannot be generated as such. 

This behavior can be verified also from Josephson potential $E_J(\chi, \gamma)$. As a criterion for determining the existence of the metastable state, we estimate the transmissivity $T_{\pi}$ from the bottom of the $\pi$-state. $T_{\pi}$ is numerically calculated as a function of $K_{\rho}$ and $K_{\sigma}$ within WKB method, and the result is shown in Fig. 4. One can see that there is a critical value of interaction strength, where $T_{\pi}$ reaches unity and the generation of $\pi$-state is prohibited completely. In the vicinity of the critical point, the tunneling probability becomes
\begin{figure}
\includegraphics[scale=0.4]{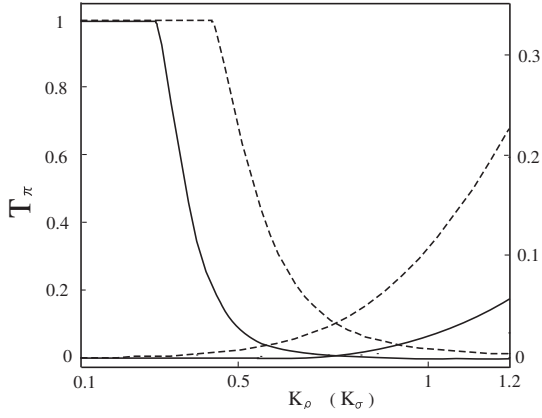}
\caption{\label{fig:epsart} The tunneling rate from the bottom of the $\pi$-state is plotted for $\gamma=0.35\pi$ (dotted line) and $\gamma=0.45\pi$ (solid line). Declining line is the one plotted as a function of $K_{\rho}$ at $K_{\sigma}=1$ (left axis), while increasing one a function of $K_{\sigma}$ at $K_{\rho}=0.8$ (right axis).}
\end{figure}
\begin{eqnarray}
T_{\pi} \sim 1-\biggr( \frac{\pi ^3}{2} \biggl)^{\frac{1}{2}}\sqrt{\frac{C}{e^2}\frac{v_F}{L}\Lambda},
\end{eqnarray}
where $C$ denotes junction capacitance. The disappearance of the metastable state is caused by the fact that the Zeeman splitting in ABSs becomes blurred due to the collective excitations arising from the interactions in LL. In practice, for repulsive interactions, the LDOS near the interfaces can be roughly estimated by
\begin{widetext}
\begin{eqnarray}
\frac{\nu(\epsilon)}{\nu(0)} \sim
\sum_{n,s} 2e^{-|\epsilon-\epsilon_{n,s}|/D} 
&\Bigl[& 
\Bigl( \frac{D \delta}{(\epsilon-\epsilon_{n,s})^2+\delta^2} \Bigr)
^{1-\zeta_{\rho}-\zeta_{\sigma}}
(\frac{v_F}{DL}) \ \Gamma(1-\zeta_{\rho}-\zeta_{\sigma})
\cos \bigl(\ 
\frac{\pi}{2}\{\zeta_{\rho}+\zeta_{\sigma}\}
\ \bigr) \nonumber \\
&+& \Bigl( 
\frac{|\epsilon-\epsilon_{n,s}|}{2D} \Bigr)^{\eta}
(\frac{v_F}{DL})^{\zeta_{\rho}+\zeta_{\sigma}} \ \Gamma(1-\eta)
\cos \bigl(\ 
\frac{\pi}{2}\{\zeta_{\rho}+\zeta_{\sigma}
-\frac{\eta}{2}\}
\ \bigr) \ 
\Bigr],
\end{eqnarray}
\end{widetext}
with the exponents $\zeta_j=(K_j^{-1}+K_{j}-2)/4$ and $\eta=(K_{\rho}^{-1}+K_{\sigma}-2)/4$. Here $\delta$ is the finite spectral weight of each ABS and $\Gamma(x)$ is Gamma function. We further fix the phase difference $\chi$ between the superconductors to zero for simplicity~\cite{rf:12}. The first term in right hand side show that the spectral weight is spread out through distribution function. In this light, the repulsive interactions can be interpreted to raise the effective temperature of the junction system. In the second one, which originates in $\Phi_{a,s}$ of Eq.(10), the repulsive interactions suppress the Andreev resonance. For attractive case ($K_{\rho}>1$), on the other hand, it becomes easier to generate the metastable state. An attractive back-scattering, in general, cannot be neglected. However, it does not change the situation qualitatively, because the attractive back-scattering tends to enhance the superconducting correlation~\cite{rf:25}. Besides, in contrast to $K_{\rho}$, small $K_{\sigma}$ enhances the critical current, and makes the $\pi$-state incident. This reveals a reflection of the symmetry of Cooper pairs in superconductors.

Finally, we mention the effect of breakdown of spin-charge separation due to the Zeeman effect. We have focused on the case where spin-charge separation is valid. This assumption may be acceptable for the system on electron doped GaAs which has small Lande's g-factor ($\sim -0.4$)~\cite{rf:4}. If, e.g., InSb quantum wires (g$ \sim -50$~\cite{rf:26}) are prepared, however, our approach needs some corrections. In such a case, the charge- and spin-density fluctuations exhibit coupled oscillation, whose coupling strength is proportional to $r=(v_{F \uparrow}-v_{F \downarrow})/(v_{F \uparrow}+v_{F \downarrow})$. Therefore, $\hat{M}$ in Eq.(12) has off-diagonal elements. This leads to different renormalization for spin-up and spin-down currents, and the ratio of renormalization factor has the following expression
\begin{eqnarray}
\frac{\Upsilon_{\uparrow}}{\Upsilon_{\downarrow}}&=&
\bigl[ 
\sinh(\frac{L}{2u_{-}L_T})/\sinh(\frac{L}{2u_{+}L_T}) 
\bigr]^{\frac{4r(u_{\rho}K_{\rho}^{-1}+u_{\sigma}K_{\sigma})}{u_+^2-u_-^2}} \nonumber \\
&\times&
\bigl[ 
\sinh(\frac{L}{2(1+r)L_T})/\sinh(\frac{L}{2(1-r)L_T}) 
\bigr]^2
,
\end{eqnarray}
where $u_{\pm}$ denotes velocity renormalization of the normal mode~\cite{rf:16}; the explicit form is given by
\begin{widetext}
\begin{eqnarray}
\!\!\!
u_{\pm}^2=\frac{1}{2} \Bigl[\ (u_{\rho}^{2}+u_{\sigma}^2+2r^2)
\pm  
\sqrt{(u_{\rho}^{2}-u_{\sigma}^2)^2+
4r^2(u_{\rho}K_{\rho}+u_{\sigma}K_{\sigma}^{-1})(u_{\rho}K_{\rho}^{-1}+u_{\sigma}K_{\sigma})} \ \Bigr].
\end{eqnarray}
\end{widetext}
Since Eq.(22) does not include high energy cut-off $D$, $\Upsilon_{\uparrow}/\Upsilon_{\downarrow}$ does not become so large. Thus the influence of the breakdown of the spin-charge separation is small, which may cause the circumstance that supercurrent is conveyed by singlet Cooper pairs.

In conclusion, current-phase relations of S-LL-S junctions are analytically reexamined using functional bosonization. We have shown that the low energy excitations in LL can play crucial roles on a critical current and a generation of $\pi$-state if LL is adiabatically connected with superconducting reservoirs. We have also examined the tunneling from the metastable $\pi$-state. When we regard the junctions as macroscopic two level systems, the quantum leakage directly affects the stability of the metastable state. The decay rate is given by $\Gamma=2Be^{-S_0-E_J(\pi,\gamma)/2}$, where $S_0=\int_0^{\pi} d\chi \sqrt{CE_J(\chi,\pi/2)/4e^2}$ is the action of symmetric potential ($B$ is the fluctuation determinant without zero mode~\cite{rf:27,rf:28}). This indicates that the macroscopic nature is inevitably related with many body correlations in 1D configuration. Our method can be easily applied to hybrid systems of superconductors with other symmetries or interacting conductor with higher dimensions~\cite{rf:21,rf:29}. Experimentally in semiconducting heterostructure, one can vary the carrier density and the effective interactions to some extent by gate voltage~\cite{rf:30}. Further tunnel junctions between 1D-2DEG have been also examined~\cite{rf:31}. It may be feasible in the future to construct the system under consideration, using S-2DEG-S junction with a gate~\cite{rf:32}.

This work is partly  supported by a Grant for The 21st Century COE Program (Physics of Self-organization Systems) at Waseda University from the Ministry of Education, Sports, Culture, Science and Technology.


\begin{thebibliography}{99}
\bibitem{rf:23} In Ref. 8, such a renormalization does not appear for perfectly transmitting regime. They consider a case that LL has strictly finite length $L$; the momentum of fluctuations takes the discrete value $n\pi/L$. Our method reproduces the result of Ref. 8 in such situation.
\end{thebibliography}

\end{document}